\documentclass[12pt,a4 paper, one side]{article} 
\usepackage{tikz}
\usepackage{amsmath,amssymb,latexsym}
\usepackage{xcolor}
\usepackage{hyperref}
\usepackage{graphicx}
\usepackage{epstopdf}
\usepackage{enumerate}
\usepackage[utf8]{inputenc}
\usepackage[T1]{fontenc}
\usepackage{amsmath}
\usepackage[leftcaption]{sidecap}
\usepackage{amssymb}
\usepackage{graphicx}
\usepackage{booktabs,dcolumn,caption}
\usepackage{rotating}
\usepackage{amsthm}
\usepackage{tikz}
\usepackage{caption}
\usepackage{subcaption}
\usepackage[most]{tcolorbox}
\begin{document}
\date{}
\title{Eternal homogeneous gravitational collapse: A comprehensive analysis from $\Theta$ parametrization}
\maketitle
\begin{center}
	\author{ Annu Jaiswal\footnote{annujais3012@gmail.com}, Rajesh Kumar\footnote{rkmath09@gmail.com}, Sudhir Kumar Srivastava\footnote{sudhirpr66@rediffmail.com}
		\\
		Department of Mathenmatics and Statistics,\\
		Deen Dayal Upadhyaya Gorakhpur University, Gorakhpur, INDIA.\\
		S.K.J.Pacif\footnote{shibesh.math@gmail.com}
		\\Centre for Cosmology and Science Popularization (CCSP), SGT University,\\
		Delhi-NCR, Gurugram 122505, Haryana, INDIA} 
\end{center}
\begin{abstract}
A new class of self-gravitating collapsing star models with perfect fluid distributions is discussed in this work. The paper has a comprehensive analysis of a homogeneous gravitational collapsing system wherein using a parametrization scheme for the expansion-scalar $(\Theta )$, the solutions of the Einstein Field Equations (EFEs) are determined independently. The background geometry for the analysis is considered to be homogeneous and isotropic represented by a Friedmann-Leimatre-Robertson-Walker (FLRW) metric and by employing the boundary conditions, we have discussed the solution in more detail. Further, all the physical and geometrical parameters are obtained in terms of Schwarzschild mass $(M)$ that makes the model significant in astrophysical applications. The singularity analysis of the
collapsing system is also discussed by the apparent-horizon, and it has seen that the homogenous gravitational collapse turns into a new kind of scenario- Eternal collapsing object.
\end{abstract}
\textbf{Keywords:} Homogeneous gravitational Collpase,  Apparent-horizon, Eternal collapsing object, Singularity, $\Theta-$ Parametrization, exact solution.\\
\textbf{MSC:} 83C05; 83F05; 83C75\\
\textbf{PACS:}  04.20.-q, 04.20.Dw, 04.20.Jb, 04.40.-b
%%%%%%%%%%%%%%%%%%%%%%%
\section{ Introduction}\label{sec1}
 The gravitational collapse, and in particular, the end state of a sufficiently massive collapsing star, is one of the most fundamental
issues in theoretical astrophysics. In a broad sense, the most of stellar objects, including stars, white-dwarfs, and neutron-stars, are products of gravitational collapsing systems. It is generally believe that the collapsing stars with ultimate masses of $4-5M_{\odot }$ (Solar mass) or greater should evolve into black holes \cite{RBB20}. It is notable that in general relativity (GR), due to the nonlinearity of Einstein's field equations, it has not been simple to find any precise solutions to the collapse difficulties by employing an equation of state (EoS) and radiation transport features of the fluids. It can be demonstrated that once the collapsing fluids are enshrouded by an event horizon, the collapse up to the central singularity is unavoidable, hence one has to comprehend the usage of the EoS up to the condition of the horizon-formation. As nuclear fusion occurs in massive stars, the gravitational collapse cannot be countered by any external heat pressure, and the star collapses to a space-time singularity \cite{SW73}\cite{RP69}. According to Penrose's Cosmic Censorship
Conjecture (CCC), the space-time singularity formed by the gravitational collapse should be hidden behind the horizon, implying that Black hole (BH) is the only feasible end state of collapse \cite{RP69}. Further, since the CCC has no proper mathematical proof and there are also various models related to the gravitational collapse of matter have been also constructed so far, where one encounters a naked singularity(NS) \cite{PS2007}.  The strong cosmic censorship conjecture states that no past extendable non-spacelike geodesic may have a positive tangent at the singularity and cannot relate it to any point on the spacetime manifold\cite{PS2007}.

\par

Thus, the final stage of a massive star's collapse is now an unresolved issue in astrophysics. At this point, it is important to remember that the Oppenheimer and Snyder model(OS) \cite{JR39} serves as the framework for the notion in the inevitable development of BH as far as the solutions are concerned. OS initiated the study of gravitational collapse with an FLRW like metric and later on several authors extended this study of gravitational collapse (see \cite{HD10}-\cite{RS18} and many more). In the following, the present investigations, describe some important configurations of the homogeneous gravitational collapse and discuss the comprehensive analysis of the model.

\par
As is widely known, in GR the motion of collapsing fluids may be characterized by the four acceleration vectors, the shear tensor, the vorticity tensor (which vanishes in our case), and the expansion scalar ($\Theta $). The space-time singularities are determined by physical,
geometrical, and kinematical characteristics that evolve throughout the gravitational collapse of a self-gravitating stellar system. The expansion scalar ($\Theta $) defines the rate of change in the elementary volume of the fluid distribution and it has a significant role in the collapsing configuration (\cite{RJ22}-\cite{h2009} and references therein). Recently, authors \cite{RJ22}-\cite{RJ23} have studied a new class of inhomogeneous gravitational collapse with uniform expansion scalar which may describes the interesting scenario of collapsing stellar systems and may also have many astrophysical consequences.

\par
The formulation of Einstein's field equations (EFEs) has allowed theoretical physicists to suggest numerous models of high-gravity astrophysical phenomena such as quasars, black holes, and other super-dense objects generated by gravitational collapse. However, the number of known exact solutions to EFEs is rather limited (or, there is not an exact solution available), which describes the realistic phenomenon of gravitational collapse. The objective of the current work is to examine the homogeneous collapse of perfect-fluid distributions from entirely new perspective and, utilizing boundary conditions, to determine the exact solution to EFEs. Since the homogeneous gravitational collapsing system requires the uniform expansion scalar $\Theta = \Theta(t)$.  We have found the exact solution of the field equations in a model-independent way. Here, we have considered a parametrization of the expansion scalar $\Theta $ i.e. a functional form of $\Theta $ as a function of time $t$, which precisely define the collapsing configurations (see figures \ref{fig1}-\ref{fig2}).

\par
In section (\ref{sec2}), the basic equations for a gravitational collapsing system is discussed in the background of FLRW spacetime metric with perfect fluid distributions and  considered the Schwarzschild metric as exterior space-time. In section (\ref{sec3}), we have introduced the mathematical parameterization of expansion scalar $\Theta $ decribing the graviational collaspe configuration.  The exact solutions of EFEs and the dynamics of models are dicussed in section (\ref{sec4}). The section (\ref{sec5}) includes the singularity analysis of a collapsing system- apparent horizon and eternal collapse scenarios. The last section (\ref{sec6}) contains the discussion and concluding remarks.

%....................................................................................

\section{General formalism for the gravitational collapse}\label{sec2} 
For the gravitational collpasing system, we consider that the space-time inside the stellar system (e.g., star) is homogeneous and isotropic, which is described by the FLRW metric 
\begin{equation}
ds_{-}^{2}=-dt^{2}+a^{2}(t)dr^{2}+R^{2} d\Omega ^{2}
 \label{eq1}
\end{equation}
where $d\Omega ^{2}=d\theta ^{2}+\sin ^{2}\theta d\phi ^{2}$ is the metric on unit 2-sphere, $a(t)$ is the scale the factor and $R=R(t,r)$ is the geometrical radius of the collapsing star given by, 
\begin{equation}
R(t,r)=r a(t)
\label{eq2}
\end{equation}
where the coordinates are taken as $x^{i}=(t,r,\theta ,\phi )$, $i=0,1,2,3$ and the fluid 4-velocity vector $V^{i}$ satisfy $V^{i}V_{i}=-1$, where $V^{i}=(1,0,0,0)$.
The energy-momentum tensor for the perfect fluid distribution is given by, 
\begin{equation}
T_{j}^{i}=diag(\rho ,p,p,p)
 \label{eq3}
\end{equation}
where $\rho $ and $p$ are the energy density and pressure respectively.

In cosmological modelling, the Hubble parameter $H=\frac{\dot{R}}{R}$ represents the expansion rate of the universe and the cosmic acceleration is described by $\dot{H}>0$ however, in the collapsing configuration $\dot{R}<0$, and the collapsing rate of the star is described by the expansion-scalar
\begin{equation}
\Theta =V_{;i}^{i}=3\frac{\dot{R}}{R}(<0)  
\label{eq5}
\end{equation}
where dot $(.)$ denotes the derivative with respect to $t$. The relation between the Hubble expansion and the expansion scalar is $\Theta =3H$ and respectively represent the expansion-rate of universe and collapsing rate of stellar system.

\par
%%%%%%%%%%%%%%%%%%%%%%%%%%%%%%%%%%%%%%%%%%%%%%%%%%%%%%%%%%%%%%%%%%%%%%%%%
The Einstein's field equations 
\begin{equation}
R_{j}^{i}-\frac{1}{2}R\delta _{j}^{i}=\mathsf{x}T_{j}^{i}  \label{efe}
\end{equation}%
for the present system yields the following two independent equations (where $\mathsf{x}=\frac{8\pi G}{c^{4}}$ ) 
\begin{equation}
\mathsf{x}\rho -\frac{1}{3}\Theta ^{2}=0  
\label{eq7}
\end{equation}

\begin{equation}
\mathsf{x}p+\frac{1}{3}\Theta ^{2}+\frac{2}{3}\dot{\Theta}=0  
\label{eq8}
\end{equation}
The mass-function $m(t,r)$ of the collapsing bodies at any moment $(t,r)$ is  described by\cite{me70}
\begin{equation}
m(t,r)=\frac{1}{2}R\left( 1+R_{,i}R_{,j}g^{ij}\right) =\frac{1}{18}\Theta
^{2}R^{3} 
 \label{eq11}
\end{equation}
where (,) denotes the partial differentiation. Also in view of Eqs. (\ref{eq7}) - (\ref{eq8}) and (\ref{eq11}), we obtain 
\begin{equation}
\dot{m}=-\frac{1}{2}\mathsf{x}p\dot{R}R^{2} 
 \label{eq12}
\end{equation}
\begin{equation}
m^{\prime }=\frac{1}{2}\mathsf{x}\rho R^{\prime 2}
 \label{eq12a}
\end{equation}

%......................................................................................

\subsection{The junction condition and the Kretschmann Curvature}\label{subsec4} 
In general relativity, Jebsen-Birkhoff's theorem states, the Schwarzschild solution is the exact solution of vacuum Einstein field equations describing the gravitational fields exterior to a spherically symmetric star, and is given by the metric,
\begin{equation}
ds^{2}_{+} = -\left(1-\frac{2M}{\xi}\right) dT^{2} + \frac{d\xi^{2}}{\left(1-
\frac{2M}{\xi}\right)} + \xi^{2} d\Omega^{2} 
\label{eq13}
\end{equation}
where $M$ represent the Newtonian mass of star (called Schwarzschild mass) and the coordinate of exterior space-time is $(T, \xi, \theta,\phi)$.
\par
The boundary hyper-surface $\Sigma$ separates the stellar system into the interior $(ds^{2}_{-})$ and the exterior $(ds^{2}_{+})$ spacetime metric. The matching of interior metric (\ref{eq1}) to the exterior Schwarzschild metric (\ref{eq13}) on the hyper-surface $\Sigma$ yield the boundary conditions \cite{NO16}
\begin{equation}
m\left(t,r\right) \overset{\Sigma}{=} M 
\label{eq14}
\end{equation}
and 
\begin{equation}
p \overset{\Sigma}{=} 0  
\label{eq15}
\end{equation}
The eq.(\ref{eq14}) shows that the mass-function $m(t,r)$ must be equal to the Schwarzschild mass $M$ on $\Sigma$ i.e., initially at $t=t_0, r=r_0$ the mass of collapsing star is described by the Schwarzschild mass $M$.
\par
The singularities are the points of the space-time, where the normal smoothness of structures of manifold break down. In other words, these are the points, where the energy density or the curvature quantities such as the scalar polynomials constructed out of the metric tensor and the Riemann tensor, diverge. One example of such a quantity is the Kretschmann scalar curvature (KS), which sometimes is called Riemann tensor squared\cite{SO02} 
\begin{equation}
\mathcal{K}=R_{i j k \delta } R^{i j k \delta }
\label{eq16}
\end{equation}
For the metric(\ref{eq1}), it gives 
\begin{equation}
\mathcal{K}=\dfrac{12\left( \dot{a}^{4}+a^{2}\ddot{a}^{2}\right) }{a^{4}}%
\text{.}  \label{eq19}
\end{equation}%
%
%%%%%%%%%%%%%%%%%%%%%%%%%%%%%%%%%%%%%%%%%%%%%%%%%%%%%%%%%%%%%%%%%%%%%%%%%%%%%%%%%%%%%

\section{$\Theta $-Parametrization}\label{sec3} 
The system of differential equations (\ref{eq7})-(\ref{eq8}) possess only two independent equations with three unknowns $a(t)$, $p(t)$
and $\rho (t)$. Therefore, it requires one more constraints for the complete determination of the solution of EFEs. In fact, a critical analysis of the solution techniques of EFEs in general relativity (or, in modified gravity theories) is the parametrization of geometrical$/$physical parameters. In literature, there are various schemes of parametrization used by researchers in cosmology (\cite{R17}-\cite{R18} and references their in).
\par
\begin{figure}[h]
	\centering
	\includegraphics{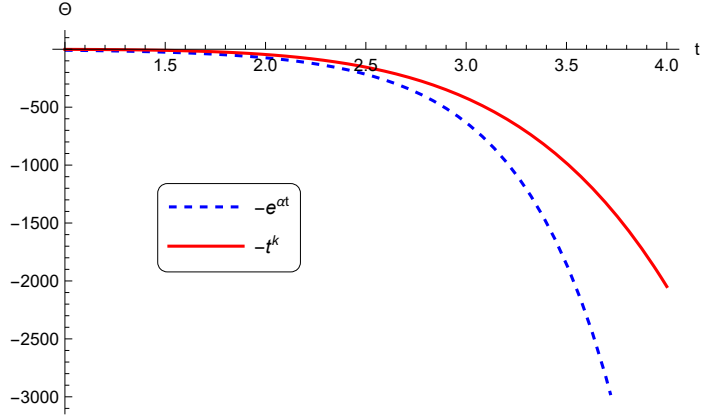}
	\caption{Collapsing configuration: variation of expansion scalar($\Theta$) with $t$  following eq.(\ref{eq20}), for $\alpha = 2.15$, $k=5.5$.}
	\label{fig1}
\end{figure}

In the evolution of stellar system (e.g., star), because of the nuclear-fusion process in the core of star, it loses its equlibrium-stage
and started to collapse under its own gravity \cite{PS2007}. During the collapsing process, the internal thermal pressure (which arises during nuclear-reaction at core) decreases and then the external pressure (which is due to the gravitational-mass of star) dominate over it. In such way the collapsing rate of star increases and hence in collapsing configuration, if one observe carefully that the collapsing rate $\Theta $ (expansion scalar) increases with $t$ following $\Theta <0$. Therefore in present study, we consider the parametrization schemes for $\Theta $ as function of t to precisely explain the notion and depict the collapsing configuration (see figures \ref{fig1} and \ref{fig2}). Recently the authors \cite{RJ22}-\cite{RJ23} have considered the various parametrization of $\Theta $ to solve the field equations for the inhomogeneous collapsing system.

\par
Here, we consider the parametrization of expansion scalar $\Theta$ as
\begin{equation}
\Theta = 
\begin{cases}
\vspace{0.4cm} -e^{\alpha t}, & \alpha \ge 0 \\ 
\vspace{0.7cm} -t^k, & k \ge 0
\end{cases}
\label{eq20}
\end{equation}
where $\alpha$ and $k$ are positive real numbers.
\par
The eq.(\ref{eq20}) is the additional constraints, has been used for the solution of EFEs (\ref{eq7})-(\ref{eq8}). In eq.(\ref{eq20}), we consider the two different $\Theta $-parametrization namely \textit{exponential} and  \textit{power law} function of $t$. In both cases, the integrating constant are determined by using the boundary condition (\ref{eq14}) and obtained the exact solutions of field equations in term of $M$. The dynamical model of collapsing phenomenon have been discussed for both cases and the comparison between both the models are described by the graphical representations ( as shown in figures (\ref{fig1})-(\ref{fig12})).

%%%%%%%%%%%%%%%%%
\section{Exact solutions and the dynamics of homogeneous gravitational Collapse}\label{sec4} 
When $\Theta =-e^{\alpha t}$ it follow from Eqs. (\ref{eq2}) and (\ref{eq5}) that 
\begin{equation}
a(t)=C_{1}e^{-\frac{{e^{t\alpha }}}{3\alpha }}  \label{eq21}
\end{equation}%
where $C_{1}$ is an integrating constant. In order to determine the value of 
$C_{1}$, we use the boundary condition(\ref{eq14}). Consider initially the
star start to collapse at ($t_{0},r_{0}$), then in view of eqs.(\ref{eq11})
and (\ref{eq21}), eq.(\ref{eq14}) gives 
\begin{equation}
r_{0}^{3}C_{1}^{3}e^{2\alpha t_{0}-\frac{1}{\alpha }e^{\alpha t_{0}}}=18M
\label{eq22}
\end{equation}%
which is equation in $C_{1}$, therefore we obtain\footnote{ Since Eq.(\ref{eq22}) is a cubic equation in $C_{1}$, hence it gives three
roots of $C_{1}$ and here we have taken only positive one.} 
\begin{equation}
C_{1}=\left( \frac{18M}{r_{0}^{3}}\right) ^{\frac{1}{3}}e^{\frac{1}{3}\left( 
\frac{1}{\alpha }e^{\alpha t}-2\alpha t_{0}\right) }  \label{eq23}
\end{equation}%
Now substituting the value of $C_{1}$ into eq. (\ref{eq21}) we obtain 
\begin{equation}
a(t)=\left( \frac{18M}{r_{0}^{3}}\right) ^{\frac{1}{3}}e^{-\frac{1}{3\alpha }%
\left( e^{\alpha t}-e^{\alpha t_{0}}+2\alpha ^{2}t_{0}\right) }  \label{eq24}
\end{equation}%
In view of eqs.(\ref{eq20}) and (\ref{eq24}) we obtain from eqs.(\ref{eq7})-(%
\ref{eq8}) that 
\begin{equation}
\mathsf{x}p=\frac{1}{3}e^{\alpha t}\left( -e^{\alpha t}+2\alpha \right) 
\label{eq25}
\end{equation}%
\begin{equation}
\mathsf{x}\rho =\frac{1}{3}e^{2\alpha t}  \label{eq26}
\end{equation}%

\begin{figure}[h]
	\centering
	\includegraphics{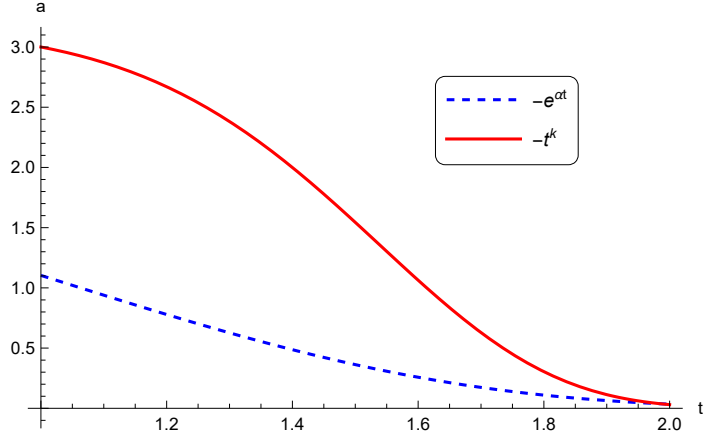}
	\caption{Collapsing configuration: variation of expansion scalar($a$) with $t$  following eq.(\ref{eq24}), where we choose $\alpha = 2.15$, $k=5.5$, $M= 1.5 M_\odot$, $r_0= 1$,$t_0=1$, and the Solar-mass $M_\odot$ is assumed to be $1$.}
	\label{fig2}
\end{figure}
\begin{figure}[h]
	\centering
	\includegraphics{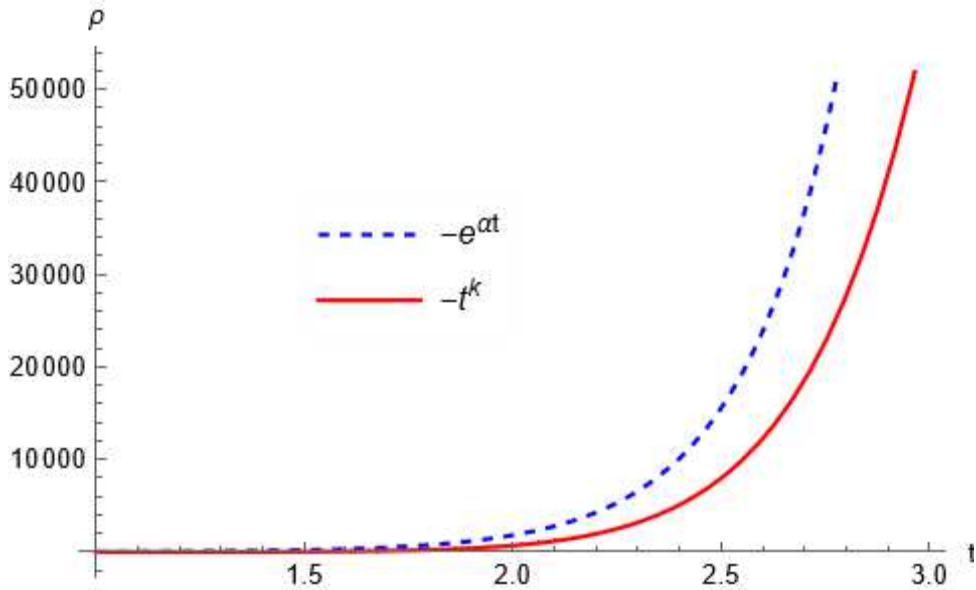}
	\caption{The variation of density($\rho$) with $t$ following eq.(\ref{eq26}),for $\alpha = 2.15$, $k=5.5$.}
	\label{fig3}
\end{figure}
%%%%%%%%%%%%%%%%%%%%%%%%%%%%
\begin{figure}[h]
	\centering
	\includegraphics{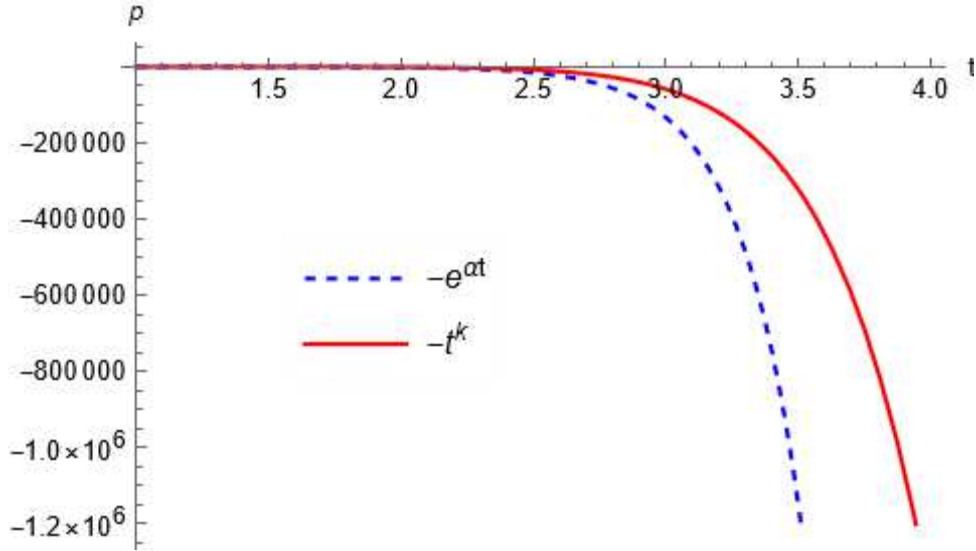}
	\caption{The variation of pressure $p$ with $t$  following eq.(\ref{eq25}),for $\alpha = 2.15$, $k=5.5$.}
	\label{fig4}
\end{figure}
\begin{figure}[h]
	\centering
	\includegraphics{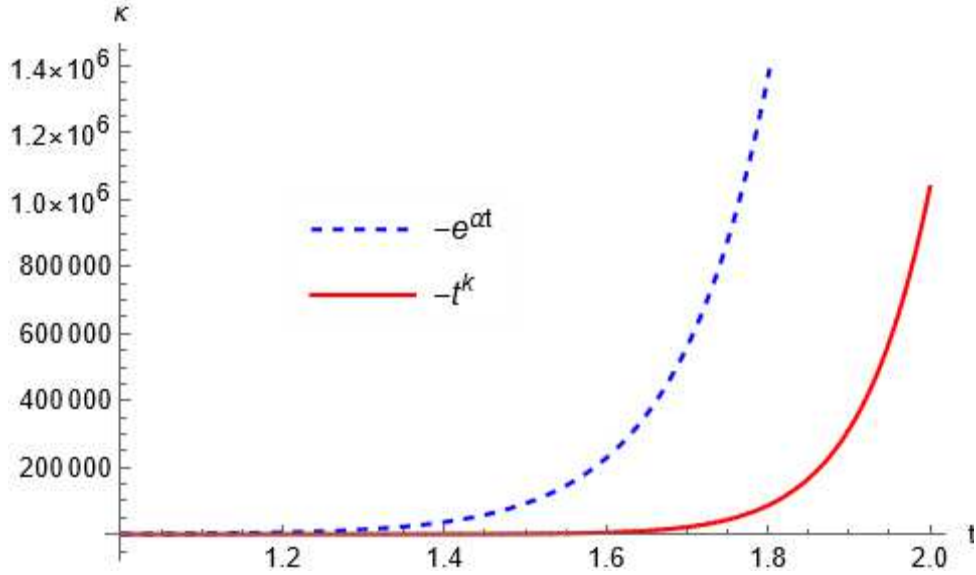}
	\caption{The variation of kretschmann scalar ($\mathcal{K}$) with $t$ following (\ref{eq28}), for $\alpha = 2.15$, $k=5.5$.}
	\label{fig5}
\end{figure}
\begin{figure}[h]
	\centering
	\includegraphics{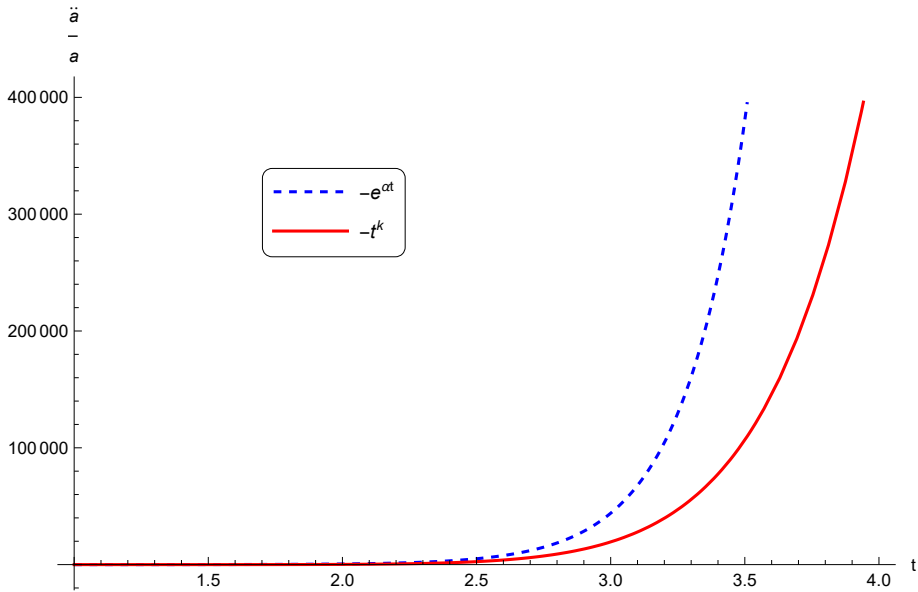}
	\caption{The variation of  acceleration($\frac{\ddot{a}}{a}$) with $t$ following (\ref{eq91}), for $\alpha = 2.15$, $k=5.5$. The graph shows the accelerating phase of collapse. }
	\label{fig6}
\end{figure}

Also from Eqs.(\ref{eq11})-(\ref{eq12a}), (\ref{eq19}) and (\ref{eq24}) we obtained the collapsing mass ($m$), rate of change of mass ($\dot{m}$), mass gradient ($m^{\prime }$) the Kretschmann curvature ($\mathcal{K}$) and the collapsing acceleration ($\frac{\ddot{a}}{a}$) 
\begin{equation}
m(t,r)=M\left( \frac{r}{r_{0}}\right) ^{3}e^{\frac{1}{\alpha }\left[ 2\alpha
^{2}\left( t-t_{0}\right) -e^{\alpha t}+e^{\alpha t_{0}}\right] }
\label{eq27}
\end{equation}%
\begin{equation}
\dot{m}=M{\left( \frac{r}{r_{0}}\right) }^{3}\left( 2\alpha -e^{\alpha
t}\right) e^{\frac{-e^{\alpha t}+2\alpha ^{2}(t-\text{t0})+e^{\alpha \text{t0%
}}}{\alpha }}  \label{eq27a}
\end{equation}
\begin{equation}
m^{\prime }=3M\frac{r^{2}}{{r_{0}}^{3}}e^{\frac{-e^{\alpha t}+2\alpha ^{2}(t-%
\text{t0})+e^{\alpha \text{t0}}}{\alpha }}  \label{eq27b}
\end{equation}

\begin{equation}
\mathcal{K} = \frac{4}{27} e^{2 \alpha t }\left( 2 e^{2 \alpha t } - 6
\alpha e^{ \alpha t} + 9 \alpha^2 \right)  \label{eq28}
\end{equation}

\begin{equation}
\frac{\ddot{a}}{a}=\frac{1}{9} e^{\alpha t} (e^{\alpha t} - 3 \alpha)
\label{eq91}
\end{equation}
%%%%%%%%

\begin{figure}[h]
	\centering
	\includegraphics{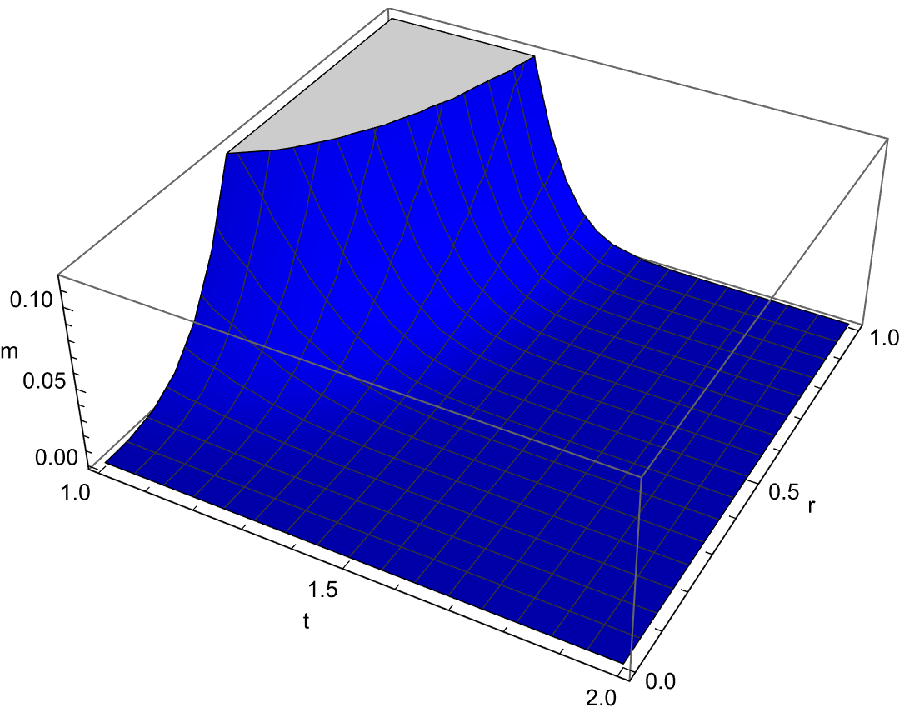}
	\caption{The variation of mass ($m$) with $t$ and $r$ for $\Theta = -e^{\alpha t}$, where we choose $\alpha = 2.15$, $M= 1.5 M_\odot$, $r_0= 1$,$t_0=1$, and the Solar-mass $M_\odot$ is assumed to be $1$.}
	\label{fig7}
\end{figure}

\begin{figure}[h]
	\centering
	\includegraphics{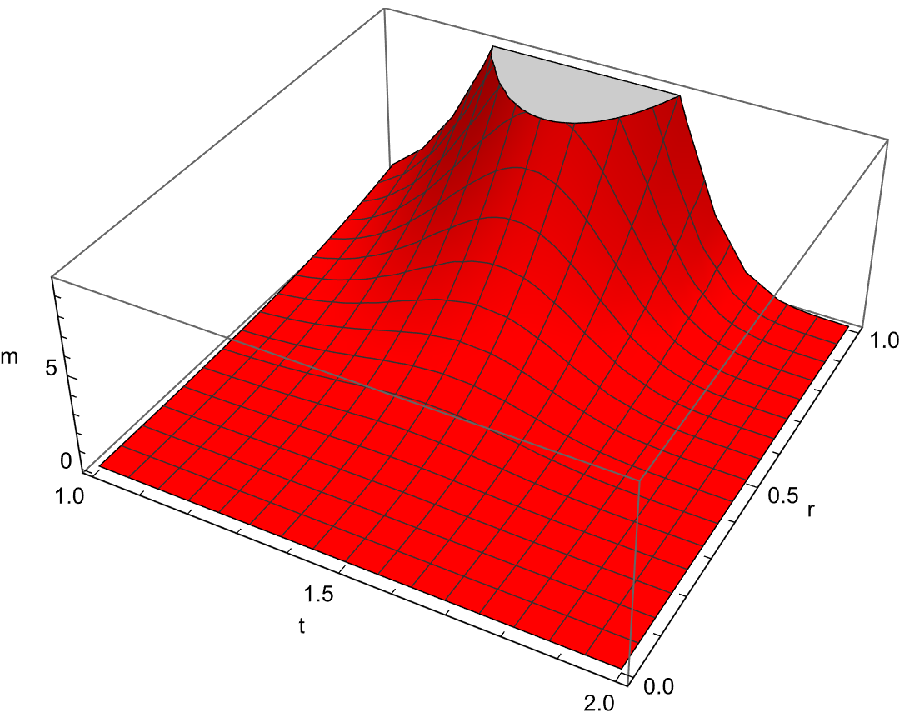}
	\caption{The variation of mass ($m$) with $t$ and $r$ for $\Theta = -t^{k}$, where we choose $k = 5.5$, $M= 1.5 M_\odot$, $r_0= 1$,$t_0=1$, and the Solar-mass $M_\odot$ is assumed to be $1$. }
	\label{fig8}
\end{figure}

\begin{figure}[h]
	\centering
	\includegraphics{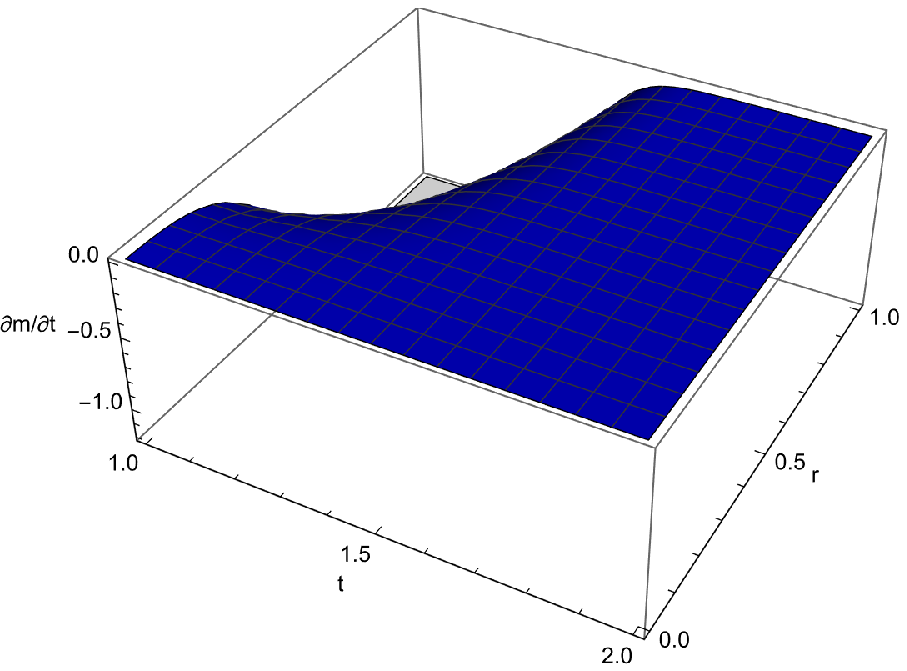}
	\caption{The behaviour of rate of change of  mass ($\dot{m}$) for $\Theta = -e^{\alpha t}$, where we choose $\alpha = 2.15$, $M= 1.5 M_\odot$, $r_0= 1$,$t_0=1$, and the Solar-mass $M_\odot$ is assumed to be $1$. }
	\label{fig9}
\end{figure}

\begin{figure}[h]
	\centering
	\includegraphics{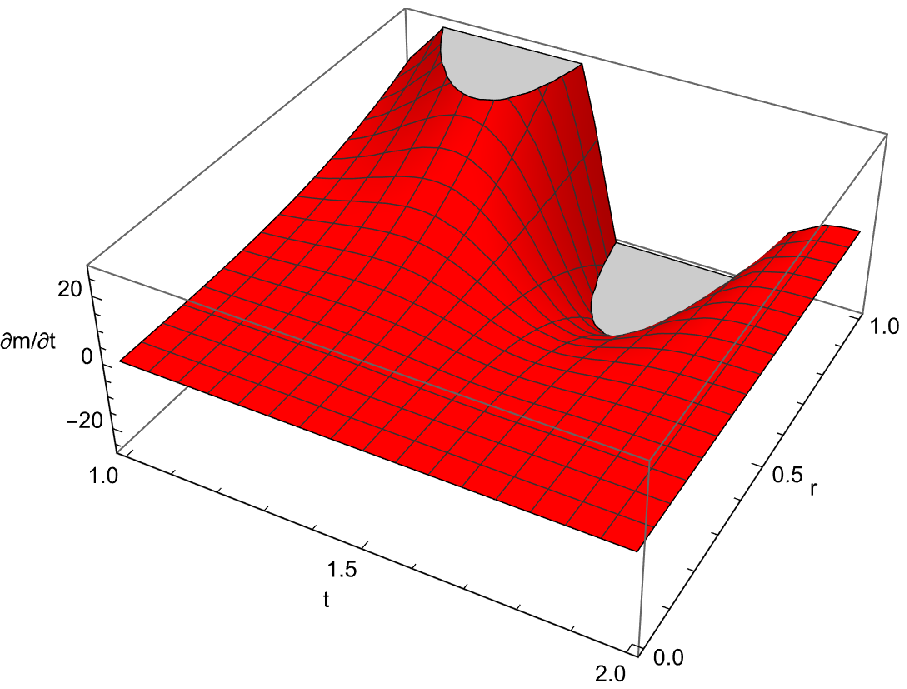}
	\caption{The behaviour of rate of change of  mass ($\dot{m}$) for $\Theta = -t^{k}$, where we choose $k = 5.5$, $M= 1.5 M_\odot$, $r_0= 1$,$t_0=1$, and the Solar-mass $M_\odot$ is assumed to be $1$. }
	\label{fig10}
\end{figure}

\begin{figure}[h]
	\centering
	\includegraphics{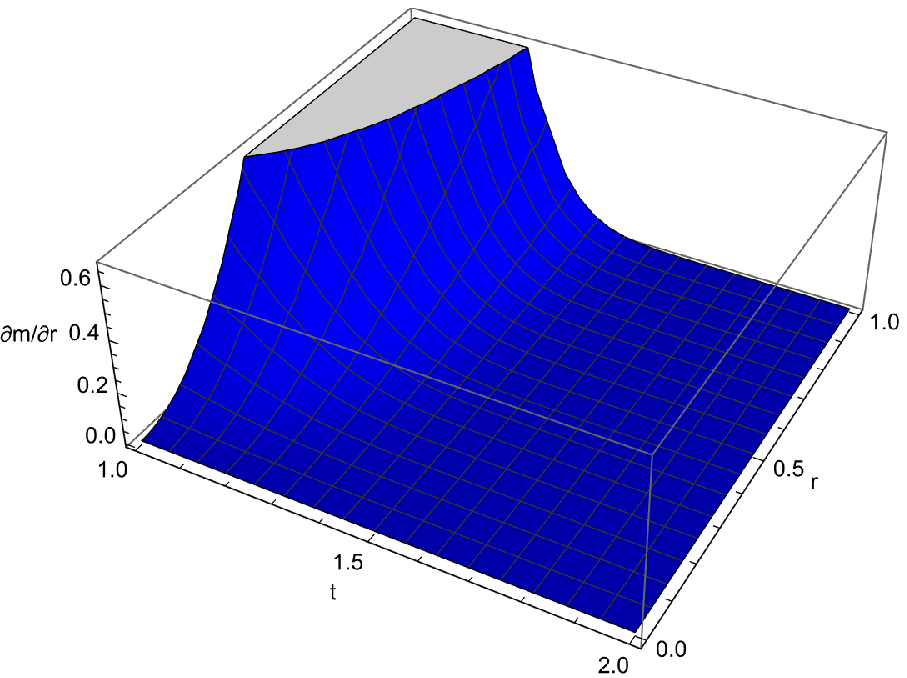}
	\caption{The behaviour of gradient of mass($m'$) for $\Theta = -e^{\alpha t}$, where we choose $\alpha = 2.15$, $M= 1.5 M_\odot$, $r_0= 1$,$t_0=1$, and the Solar-mass $M_\odot$ is assumed to be $1$.}
	\label{fig11}
\end{figure}

\begin{figure}[h]
	\centering
	\includegraphics{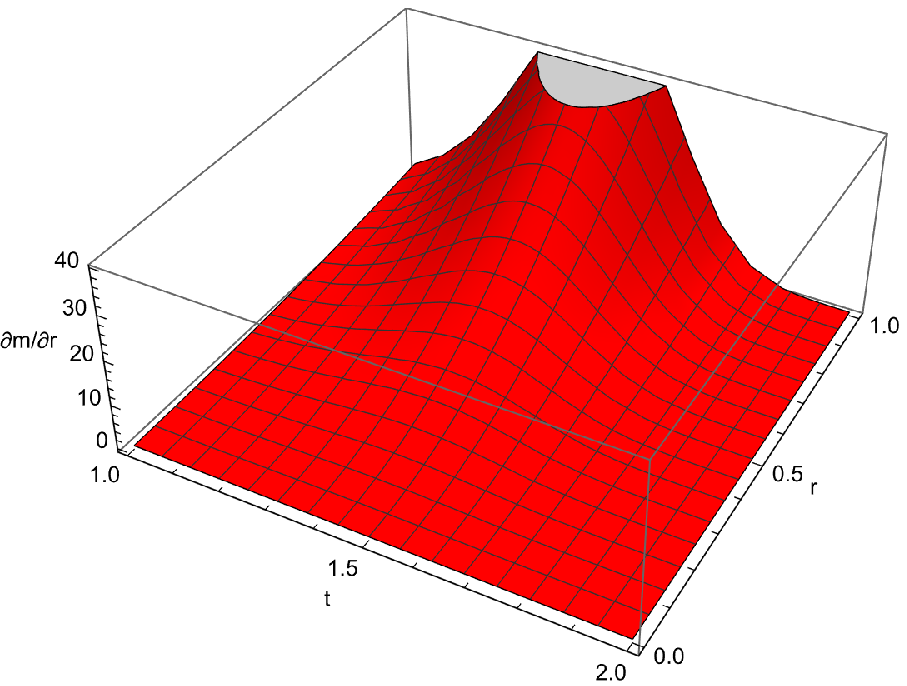}
	\caption{The behaviour of gradient of mass($m'$) for $\Theta = -t^{k}$, where we choose $k = 5.5$, $M= 1.5 M_\odot$, $r_0= 1$,$t_0=1$, and the Solar-mass $M_\odot$ is assumed to be $1$.}
	\label{fig12}
\end{figure}

Similarly, for the power law parametrization $\Theta = -t^k $, we have obtained the values of geometrical and physical quantities which are
summarize in Table (\ref{table1}), (\ref{table2}) and (\ref{table3}). The dynamical behaviour of $\rho, p, \mathcal{K}, \frac{\ddot{a}}{a}$ and $m$ are shown in figures (\ref{fig1})-(\ref{fig12}).
\captionsetup{labelsep=newline,
	singlelinecheck=false,
	skip=0.333\baselineskip} \newcolumntype{d}[1]{D{.}{.}{#1}} 
% "decimal" column type
\renewcommand{\ast}{{}^{\textstyle *}} % for raised "asterisks"
\begin{sidewaystable}[h]
		\caption{scale factor($a$), density($\rho$) and pressure($p$) for both parametrization of $\Theta$}
			\label {table1}
		\begin{tabular}{p{2.4cm}p{6.3cm}p{3.5cm}p{4cm}}
			\toprule
			$\Theta$ & a(t) &$\rho$& $p$ \\
			\midrule
			$-e^{\alpha t}$ & $ \left(\frac{18 M}{r^3_0}\right)^{\frac{1}{3}} e^{-\frac{\left( e^{\alpha t} - e^{\alpha t_0} + 2 \alpha^2 t_0\right)}{{3\alpha}}}$ & $	\frac{1}{3\mathsf{x}} e^{2\alpha t}$ & $-\frac{1}{3\mathsf{x}}(e^{\alpha t} \left(e^{\alpha t} - 2\alpha\right))$  \\ \addlinespace \addlinespace \addlinespace
			$-t^{k}$ & $\left(\frac{18 M}{r_{0}^3}\right)^{(1/3)} t_0^{(-2k/3)} e^{\frac{1}{3(1+k)}[ -t^{1+k}  + t_{0}^{1+k}]}$ & $\frac{1}{3{\mathsf{x}}}t^{2k}$ & $\frac{-1}{3{\mathsf{x}}} (-2k t^{-1+k} + t^{2k})$\\ \addlinespace \addlinespace \addlinespace
			\bottomrule

		\end{tabular}
	\end{sidewaystable}
\captionsetup{labelsep=newline,
	singlelinecheck=false,
	skip=0.333\baselineskip} \newcolumntype{d}[1]{D{.}{.}{#1}} 
% "decimal" column type
\renewcommand{\ast}{{}^{\textstyle *}} % for raised "asterisks"
\begin{sidewaystable}[h]
	\caption{Kretschmann curvature ($\mathcal{K}$) and collapsing acceleartion($\frac{\ddot{a}}{a}$) for both parametrization of $\Theta$}
	\label {table2}
	\begin{tabular}{p{3.9cm}p{6.9cm}p{3.5cm}p{2cm}}
		\toprule
		$\Theta$ & $\mathcal{K}$ &$\frac{\ddot{a}}{a}$ & \\
		\midrule
		$-e^{\alpha t}$ &  $ \frac{4}{27}e^{2 \alpha t }(2 e^{2  \alpha t } - 6 e^{ \alpha t} \alpha  + 9 \alpha^2)$ &  $\frac{1}{9} e^{\alpha t} (e^{\alpha t} - 3 \alpha)$ &\\ \rule{0pt}{10ex}
		$-t^{k}$  & $\frac{4}{27}t^{-2+2k}(2t^{2+2k}-6t^{1+k}k +9k^2)$ & $\frac{1}{9} t^{-1+k} (-3k + t^{1+k})$ &\\ \addlinespace \addlinespace 
		\bottomrule

	\end{tabular}
\end{sidewaystable}
\captionsetup{labelsep=newline,
	singlelinecheck=false,
	skip=0.333\baselineskip} \newcolumntype{d}[1]{D{.}{.}{#1}} 
% "decimal" column type
\renewcommand{\ast}{{}^{\textstyle *}} % for raised "asterisks"
\begin{sidewaystable}[h]
	\caption{mass($m$), rate of change of mass($\dot{m}$) and mass-gradient($m'$) for both parametrization of $\Theta$}
	\label {table3}
	\begin{tabular}{p{2.0cm}p{6.5cm}p{6cm}p{5.3cm}p{0.6cm}}
		\toprule
		$\Theta$ &m &$\dot{m}$&$m^{'}$ &\\
		\midrule
		$-e^{\alpha t}$&$M(\frac{r}{r_0})^3 e^{\frac{-e^{\alpha t}+ e^{\alpha t_0 }+2 \alpha^{2}(t-t_0)}{\alpha}}$&$-M(\frac{r}{r_0})^3(e^{\alpha t}-2\alpha)\newline e^{\frac{-e^{\alpha t}+ e^{\alpha t_0 }+2 \alpha^{2}(t - t_0)}{\alpha}}$&$\frac{3 e^\frac{{-e^{\alpha t} + e^{\alpha t_{0}} + 2(t-t_{0})\alpha^2}}{\alpha}M r^2} {r^3_{0}}$& \\ \rule{0pt}{10ex}
		$-t^{k}$&$M (\frac{r}{r_0})^{3} (t-\frac{1}{t_0})^{2k} e^{\frac{-t^{1+k} + t_0^{1+k}}{1+k} }$&$M (\frac{r}{r_0})^{3} t^{2k-1} t^{-2k}_{0}(2k- t^{1+k})\newline  e^{\frac{-t^{1+k} + t_0^{1+k}}{1+k} }$&$\frac{3 e^{\frac{-t^{1+k} + t^{1+k}_{0}}{1+k}M r^2 t^{2k} t^{-2k}_{0}}}{r^3_{0}} $&\\ \addlinespace \addlinespace
		\bottomrule

	\end{tabular}
\end{sidewaystable}


%%%%%%%%%%%%%%%%%%%%%%%%%%%%%%%%%%%%%%%%%%%%%

\section{Singularity Analysis: Apparent Horizon and Eternal collapse}\label{sec5}
The development of a space-time singularity, which is characterized by the divergence of the curvature $\mathcal{K}$ and the energy density $\rho $, is generally outcome of the gravitational collapse (GC) of self-gravitating systems. The possible outcomes of GC in terms of either a BH or NS are then specified by the occurrence of trapped-surfaces developing in the space-time as the gravitational-collapse progresses. Initially, when object starts collapse under the effect of its own-gravity, no portions of the space-time are trapped but as critical high density are reached, the trapped-surfaces form and apparent-horizon region develops in the space-time \cite{S2}-\cite{S3}. The singularity can be causally connected to or disconnected from the outside universe, depending on the sequence of trapped-surface formation as the collapse evolves, as it has been observed that the apparent-horizon typically develops between the time of singularity's formation and the time at which it meets the outer Schwarzschild event horizon \cite{S2}- \cite{S4}.

\par
In the BH scenario, the apparent-horizon develops at a stage earlier than the singularity-formation. As the singularity occurs, the event horizon on the outside space-time completely covers the final stages of collapse, while the apparent horizon within the matter develops from the outside shell to the singularity at the moment of its formation \cite{S5}-\cite{S10}. In NS, the trapped-surface develops in the cloud's centre at the moment the singularity forms, and the apparent-horizon then moves outward to meet the event-horizon at the boundary later than the singularity formation \cite{S10}.

For the present FLRW space-time metric(\ref{eq1}), the apparent-horizon is described by 
\begin{equation}
R_{,i} R_{,j} g^{ij} = \left(r \dot{a}\right)^2 - 1 = 0  \label{eq39}
\end{equation}
where the comma (,) denotes the partial derivatives and $R = r a$.
\par
Since the present study concern of the singularity formation due to the gravitational collpase of star, let us assume that initially when $(t_{0},r_{0})$ the star is not trapped i.e., 
\begin{equation}
R_{,i} R_{,j} g^{ij}|_{(t_{0},r_{0})} = \left( r_{0}^2 \dot{a}^2(t_{0})
\right) - 1 < 0  
\label{eq399}
\end{equation}
Further, let us assume  at $(t_{AH},r_{AH})$ the whole star collapses inside the apparent-horizon, then it follows from eq.(\ref{eq39}) that 
\begin{equation}
\dot{R}^2(t_{AH},r_{AH})-1= r^2_{AH} \dot{a}^2(t_{AH})-1 = 0  
\label{eq400}
\end{equation}
In the first case $\Theta = -e^{\alpha t}$, by using eq.(\ref{eq24}) into (\ref{eq400}) we obtain the time formation of apparent horizon 
\begin{equation}
t_{AH} = \frac{1}{6\alpha^2} \left[-2e^{\alpha t_{0}}+4\alpha^2 t_0 +
3\alpha Log \left(\frac{(\frac{3}{2})^{\frac{2}{3}}r_0^2}{M^{\frac{3}{2}}
r_{AH}^2}\right) - 6\alpha \mathcal{W(X)} \right]  
\label{eq42}
\end{equation}
where $\mathcal{X}= -\frac{1}{(18M)^{\frac{1}{3}}} \left( \frac{r_0}{\alpha r_{AH}} \right) e^{-\frac{1}{3\alpha}e^{\alpha t_0}+\frac{2}{3} \alpha t_0} $ and $\mathcal{W}$ denotes the \textit{Lambert W-function}\footnote{In mathematics, the Lambert W-function is also known as ProductLog function or, \textit{Omega function}\cite{BD2000}} defined by the equation $\mathcal{W(X)} e^{\mathcal{X}}=\mathcal{X}$ (see \cite{BD2000} for more details). For $\mathcal{X} \geq -\frac{1}{e} = -0.3679$, $\mathcal{W(X)}$ gives the real values. Here we can see that the value of $\mathcal{X}$ depends on $\alpha, t_0, r_0, r_{AH}$ and $M$. For $M \geq 1.4 M_\odot$, $\alpha=2.15, (t_0, r_0) = (1,1), r_{AH} = 0.02$, we get the value of $\mathcal{X} = -7.81648 \times 10^{-10}$, where $M_\odot = 1.989 \times 10^{30} Kg$ is the Solar mass and $1.4 M_\odot$ is the Chandrashekhar-limit mass. In such a way we see that eq.(\ref{eq42}) gives the finite real value of $t_{AH}$(time
formation of apparent horizon\footnote{One can also check that for any arbitrary values of $\alpha, r_{AH}$ and initial coordinate $(t_0, r_0)$, $\mathcal{X} > -0.3679$ and $t_{AH}$ has finite real value for all $M \geq 1.4 M_\odot$.})

\par
The geometrical radius of apparent-horizon surface is 
\begin{equation}
R_{AH} = r_{AH} a(t_{AH}) = (18M)^{\frac{1}{3}} \frac{r_{AH}}{r_0} e^{{\frac{%
1}{3 \alpha}( e^{\alpha t_0} - e^{\alpha t_{AH}} -2\alpha^2 t_0)}}
\label{eq401}
\end{equation}
The total contribution of collapsing star to the mass of apparent-horizon region is 
\begin{equation}
M_{AH} = m|_{(t_{AH},r_{AH})} = \left(\frac{9 M}{4}\right)^{\frac{1}{3}} 
\frac{r_{AH}}{r_0} e^{\frac{-1}{6 \alpha} \left(\mathcal{M} (\frac{12}{M})^{%
\frac{1}{3}} \frac{r_{AH}}{r_0} +4 \alpha^2 t_0 -2e^{\alpha t_0}\right) }
\label{eq402}
\end{equation}

where $\mathcal{M} = e^{ \frac{2}{3} \alpha t_0 - \frac{1}{3\alpha} e^{\alpha t_0} } -\mathcal{W(X)}$

Analogously, for the second case $\Theta = -t^k$, we obtained the value of $t_{AH}$, $R_{AH}$ and $M_{AH}$ which are summarized in Table(\ref{table4}).
\par
\textbf{Occurance of Eternal collapse phenamenon}
\par
\begin{figure}[h]
	\centering
	\includegraphics[scale=0.6]{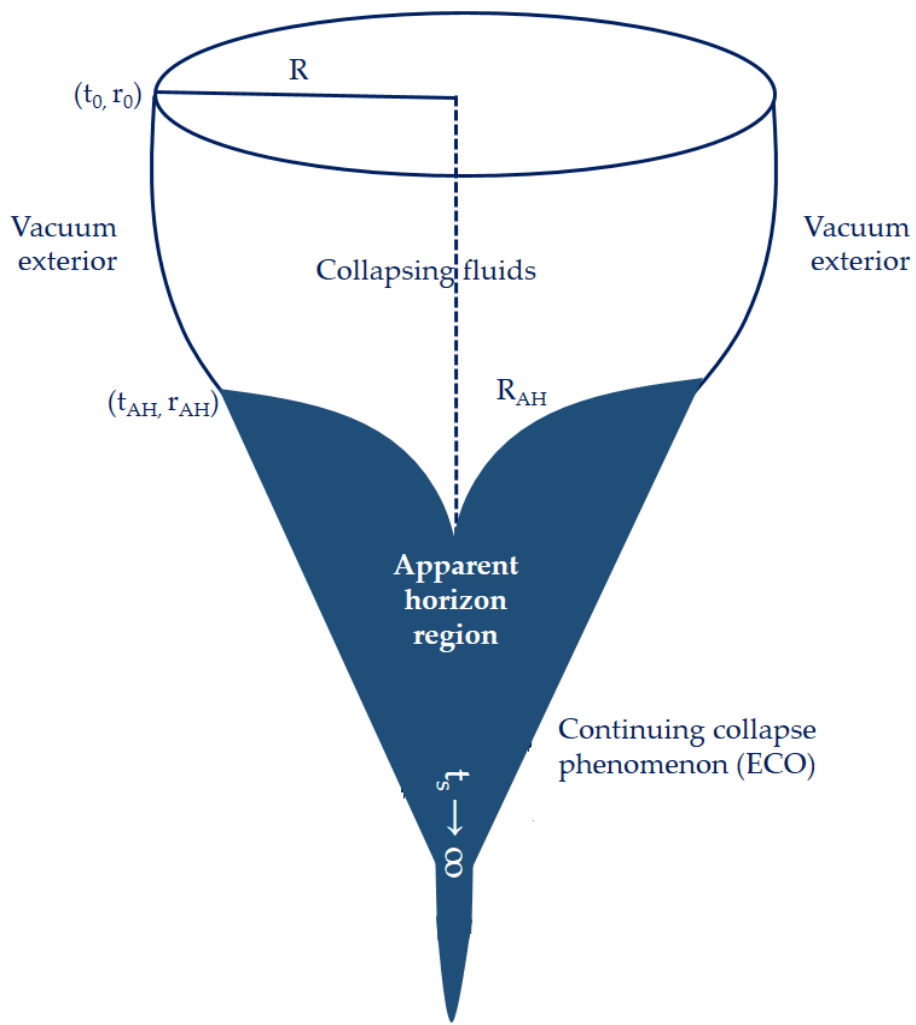}
	\caption{Eternal collapsing object: The figure shows the continuous collapsing to attain the singularity at infinity}
	\label{fig13}
\end{figure}

Assuming that the star begin to collapse at the moment $t=t_{0}$ where condition (\ref{eq399}) hold i.e., the star is not initially trapped. From Table(\ref{table1}) and Table(\ref{table2}) we can see that $\rho \rightarrow \infty $, $\mathcal{K}\rightarrow \infty $ as $t=t_{s}\rightarrow \infty $, in other words the energy density and Kretschmann curvature diverge at infinite commoving time and hence star tend to collapse for infinite duration in order to attain the space-time singularity. Further, it can be seen from Table(\ref{table4}) that apparent-horizons form in a finite commoving time $t_{AH}$ much earlier than the time of collapse ($t_{s}=\infty $). Thus the singularity is not naked because before it is formed an apparent-horizon is already formed at $t_{AH}$ (see Table(\ref{table4})). Also from Table(\ref{table3}), we see that mass vanishes at the time of collapse $(t_{s}\rightarrow \infty $. Because both the final BH mass and the commoving time for its formation are finite, i.e., the BH candidate must be formed during gravitational collapse with finite
mass in a finite time rather than $m\rightarrow 0$ and $t_{s}\rightarrow infty $ and therefore, the BH is also not formed here. Also, wee that the acceleration is continuously increases showing accelerating phase of gravitational collapse(as can be seen in figure \ref{fig6}). Hence, we
conclude that as homogeneous gravitating system tend to collapse for infinite commoving time in order to attain the singular state $(\rho
\rightarrow \infty ,\mathcal{K}\rightarrow \infty )$ and therefore may be called "Eternal Collapsing Object (ECO)" \cite{ne34}\cite{ne34a}. The whole scenario of eternal collapsing star is shown in figure(\ref{fig13})

%%%%%%%%%%%%%%%%%%%%%%%%%%%%%%%%%%%%%%%%%%%%%%%%%%%%%%%%%%%%%%%%%%%%%%%%
%%%%%%%%%%%%%%%%%%%%%%%%%%%%%%%%%%%%%%%%%%%%%%%%%%%%%%%%%%%%%%%%%%%%%%%%
\captionsetup{labelsep=newline,
 	singlelinecheck=false,
 	skip=0.333\baselineskip} \newcolumntype{d}[1]{D{.}{.}{#1}} 
% "decimal" column type
\renewcommand{\ast}{{}^{\textstyle *}} % for raised "asterisks"
\renewcommand\arraystretch{1.5} 
\begin{sidewaystable}[h]
 	\caption{Apparent horizon: time ($t_{AH}$),surface radius($R_{AH}$) and mass($M_{AH}$) for both parametrization of $\Theta$}
 	\label {table4}
 	\begin{tabular}{p{1.1cm}p{8.9cm}p{7.0cm}p{2.0cm}}
 	\hline
 	$\Theta$ &   $t_{AH}$  &  $R_{AH}$   &  $M_{AH}$  \\
 	\hline
 	$-e^{\alpha t}$ & $\frac{1}{6\alpha^2} [-2e^{\alpha t_{0}}+4\alpha^2 t_0 +\newline 3\alpha Log(\frac{(\frac{3}{2})^{\frac{2}{3}}r_0^2}{M^{\frac{3}{2}} r_{AH}^2})
 	-  6\alpha  \mathcal{W(X)}]$ & $(18M)^{\frac{1}{3}} \frac{r_{AH}}{r_0} \newline e^{{\frac{1}{3 \alpha}( e^{\alpha t_0} - e^{\alpha t_{AH}} -2\alpha^2 t_0)}}  $ & $(\frac{9 M}{4})^{\frac{1}{3}} \frac{r_{AH}}{r_0} e^{\frac{-1}{6 \alpha}( \mathcal{M} (\frac{12}{M})^{\frac{1}{3}} \frac{r_{AH}}{r_0} +4 \alpha^2 t_0 -2e^{\alpha t_0}) }$ \\ \rule{0pt}{10ex}
 	$-t^{k}$ & $ \frac{(\frac{2M}{3})^{\frac{2}{3}} r_{AH}^2 t_0^{\frac{-4k}{3} t_{AH}^{2k} e^{\frac{2(t_0^{k+1}-t_{AH}^{k+1})}{3(k+1)}}}  }{r_0^2} =1$\footnote{This equation gives the finite vlaue of $t_{AH}$ } & $	2^{\frac{1}{3}}  3^{\frac{2}{3}}e^{\frac{-t^{1+k} + t_{0}^{1+K}}{3(1+k)}} (\frac{M}{r_{0}^3})^{\frac{1}{3}} r_{AH} t_{0}^{\frac{-2k}{3}}$ & $\frac{3}{2} t_{AH}^{-k}$\\ \addlinespace 
 	\hline 
 	\end{tabular}
 \end{sidewaystable}
%%%%%%%%%%%%%%%%%%%%%%%%%%%%%%%%%%%%%%%%%%%%%%%%%%%%%%%%%%%%%

\section{Discussion and Concluding Remarks}\label{sec6} 
The objective of this work is to discuss the physical process during homogeneous gravitational collapsing phase of a stellar systems and
its final state. Studies of exact solutions and the singularities formation play an crucial role in general relativity, even in the current context. There is no single method preferred for finding solutions to the EFEs. Although many exact solutions exist, very few pose physically interesting results and that too in a very restricted scenario. From this point of view, the present investigations is performed regarding the gravitational collapse  and a comprehensive analysis of singularity formation in the background of homogeneous and isotropic FLRW geometry. Here, we have considered two different parameterization of $\Theta $ as functions of $t$, namely, exponential ($-e^{\alpha t}$) and power law ($-t^{k}$), which precisely describes the collapsing process of stellar system (as shown in figure (\ref{fig1})). In addition, we have applied the boundary condition to explicitly obtain the exact solutions in terms of mass $M$.

\par
A systematic discussion is presented, assuming the parameterization (\ref{eq20}) and obtained the exact solutions in the explicit form (as shown in tables (\ref{table1}) and (\ref{table2})). We have also discussed the collapsing mass($m$), mass-rate ($\dot{m}$) and mass-gradient ($m^{\prime }$) during the collapsing configuration as can be seen in Table (\ref{table3}). It is observed that the scale factor $a$ is decreasing whereas the pressure $p(<0)$, density $\rho $ and Kretschman curvature $k$ are continuously increasing during the collapsing process (as shown in figures (\ref{fig2})-(\ref{fig5})). Also, from eq.(\ref{eq91}) and Table-\ref{table2}, we have shown that our models represents an accelerating phase of collapse (as shown in fig-\ref{fig6}) and hence the ever-increasing Kretschmann curvature and density, which tend to extend physical space-time to an infinite extent, the collapse of sufficiently massive objects may continue forever. In section (\ref{sec5}), the formation of apparent-horizon is discussed and it turns out that the apparent-horizon develop before the singularity-formation $t_{AH}<t_{s}$ (see Table-\ref{table4}). We have also obtained the geometric radius $R_{AH}$ and finite mass $M_{AH}$ of the apparent-horizon surface which are summarize in Table-\ref{table4}. From Table-\ref{table3}, we see that the mass $m(t,r)$ decreases during such collapsing configuration (as shown in figures (\ref{fig7})-(\ref{fig8})). The ECO is massive and continuing collapse which try to attain the singular state in an infinite time and its mass would be $m\rightarrow 0$ as $t_{s}\rightarrow \infty $. It can be seen that $\dot{m}$ continuously decreases which shows the loss of mass with time and the gradient of mass ($m^{\prime }$) is also decreases continuously in both cases $\Theta =-e^{\alpha t}$ and $\Theta =-t^{k}$ as shown in figures \ref{fig9}-\ref{fig12}.

\par

 For the definite conclusion about the singularity formation, we have examined the final state of the collapsing process by comparing the time of singularity formation ($t_{s}$) and the time of the apparent horizon formation ($t_{AH}$). From the behavior of Kretschmann curvature and the energy density are seemed to be divergent at $t=t_{s}\rightarrow \infty $  and it is seen that the singularity occurs as a result of the
 gravitational-collapse in an infinite time. Therefore, it is probable that the gravitational collapse may find either quasi-stable ultra-compact or even continuing collapse configurations, subject to the as of yet unknown behavior of the high density and as of yet undefined plausible phase transitions of collapsing matter under such conditions. Thus, such collapsing objects are more massive and more compact have been treated as Eternal Collapsing Object(ECO) \cite{ne34}\cite{ne34a}.

\par
Additionally, it should be noted that our work has supported the conclusions of Misner \cite{mw69}, who claimed that even though we are well aware of the possibility of general relativity failing as we approach the ultra-high-density region, we should still consider the predictions of general relativity in this regime since they may provide some evidence as to what to expect from a more general theory of gravity that performs in this regime. Our model here present very much of physically realistic stellar systems because the obtained solutions with all the physical and geometrical parameters are in terms of Mass $M$ of star and hence it may be explored towards astrophysically more realistic stellar objects. Although, we have not examined here the comparison of astrophysics observations of stars with our result but we plan to take up these studies in the future works. The advantage of our model, however, is that we have presented here a fully consistent general relativistic model to describes the collapsing scenarios  of stellar objects of known mass $M$ and radius $R$ and the idea can also be  explored in modified gravity models too and is deferred to our future investigations.

%%%%%%%%%%%%%%%%%%%%%%%%%%%%%%%%%%%%%%%%%%%%%%%%%%%%%%%

\textbf{Acknowledgment:} The authors AJ, RK and SKS are acknowledge to the Council of Science and Technology, UP, India vide letter no. CST/D-2289. 
%%%..................................................................................................................


\begin{thebibliography}{99}
\bibitem{RBB20} Raychaudhuri, A. K., Banerji, S., and Banerjee, A. (2003). \textit{General relativity, astrophysics, and cosmology}. Springer Science \& Business Media.

\bibitem{SW73} Hawking, S. W., and Ellis, G. F. (2023). \textit{The large scale structure of space-time}. Cambridge university press.

\bibitem{RP69} Penrose, R. (1969).\textit{ Gravitational collapse: The role of general relativity}. Nuovo Cimento Rivista Serie, 1, 252.

\bibitem{PS2007} Joshi, P. S. (2007). \textit{Gravitational collapse and spacetime singularities} (Vol. 2). Cambridge: Cambridge University Press.


\bibitem{JR39} Oppenheimer, J. R., \& Snyder, H. (1939). \textit{On continued gravitational contraction}. Physical Review, 56(5), 455.

\bibitem{HD10} Herrera, L., Di Prisco, A., Ospino, J., \& Carot, J. (2010). \textit{Lemaitre-Tolman-Bondi dust spacetimes: Symmetry properties and some extensions to the dissipative case}. Physical Review D, 82(2), 024021.

\bibitem{M72} Misra, R. M., \& Srivastava, D. C. (1972). \textit{Gravitational Collapse of Homogeneous Spheres}. Nature Physical Science, 238(86), 116-117.

\bibitem{RS18} Kumar, R., \& Srivastava, S. K. (2018).\textit{ Expansion-free self-gravitating dust dissipative fluids}. General Relativity and Gravitation, 50, 1-16.

\bibitem{RJ22} Kumar, R., \& Jaiswal, A. (2022). \textit{A new class of spherically symmetric gravitational collapse}. Theoretical and Mathematical Physics, 211(1), 558-566.

\bibitem{RJ23} Jaiswal, A., Srivastava, S. K., \& Kumar, R. (2023). \textit{Dynamics of uniformally collapsing system and the horizon formation}. International Journal of Geometric Methods in Modern Physics, 2350114.

\bibitem{h2009} Herrera, L., Santos, N.O. \& Wang, A. (2008). \textit{Shearing expansion-free spherically symmetric anisotropic fluid evolution}. Physical Review D, 78(8),084026.


\bibitem{me70} Cahill, M. E., \& McVittie, G. C. (1970). \textit{Spherical Symmetry and Mass‐Energy in General Relativity. I. General Theory}. Journal of Mathematical Physics, 11(4), 1382-1391.

\bibitem{NO16} Santos, N. O. (1985). \textit{Non-adiabatic radiating collapse}. Monthly Notices of the Royal Astronomical Society (ISSN 0035-8711), vol. 216, Sept. 15, 1985, p. 403-410. Research supported by the Coordenacao do Aperfeicoamento do Pessoal de Ensino Superior., 216, 403-410.

\bibitem{SO02} Cherubini, C., Bini, D., Capozziello, S., \& Ruffini, R. (2002). \textit{Second order scalar invariants of the Riemann tensor: applications to black hole spacetimes}. International Journal of Modern Physics D, 11(06), 827-841.

\bibitem{R17} Pacif, S. K. J., Myrzakulov, R., \& Myrzakul, S. (2017).\textit{ Reconstruction of cosmic history from a simple parametrization of H}. International Journal of Geometric Methods in Modern Physics, 14(07), 1750111.

\bibitem{R18} Pacif, S. K. J. (2020). \textit{Dark energy models from a parametrization of H: a comprehensive analysis and observational constraints}. The European Physical Journal Plus, 135(10), 1-34.

\bibitem{S2} Anninos, P., Bernstein, D., Brandt, S. R., Hobill, D., Seidel, E., \& Smarr, L. (1994). \textit{Dynamics of black hole apparent horizons}. Physical Review D, 50(6), 3801.

\bibitem{S3} Bizon, P., Malec, E., \& O'Murchadha, N. (1988). \textit{Trapped surfaces in spherical stars}. Physical review letters, 61(10), 1147.

\bibitem{S4} Ellis, G. F. (2003). \textit{Closed trapped surfaces in cosmology}. General Relativity and Gravitation, 35, 1309-1319.

\bibitem{S5} Weinberg, S. (1972). \textit{Gravitation and cosmology: principles and applications of the general theory of relativity}.

\bibitem{S10} Bhattacharjee, S., Saha, S., \& Chakraborty, S. (2018). \textit{Does particle creation mechanism favour formation of black hole or naked singularity $?$}. The European Physical Journal C, 78, 1-18.

\bibitem{BD2000} Barry, D. A., Parlange, J. Y., Li, L., Prommer, H., Cunningham, C. J., \& Stagnitti, F. (2000). \textit{Analytical approximations for real values of the Lambert W-function}. Mathematics and Computers in Simulation, 53(1-2), 95-103.

\bibitem{ne34} Mitra, A. (2006). \textit{Radiation pressure supported stars in Einstein gravity: eternally collapsing objects}. Monthly Notices of the Royal Astronomical Society, 369(1), 492-496.

\bibitem{ne34a} Mitra, A., \& Glendenning, N. K. (2010).\textit{ Likely formation of general relativistic radiation pressure supported stars or ‘eternally collapsing objects’}. Monthly Notices of the Royal Astronomical Society: Letters, 404(1), L50-L54.

\bibitem{mw69} Misner, C. W. (1969). \textit{Absolute zero of time.} Physical Review, 186(5), 1328.

%%%%%%%%%%%%%%%%%%%%%%%%%%%%%%%%%%%%%%%%%%%%%%
\end{thebibliography}
\end{document}